\documentclass[
    ,final            
  ]
  {aipproc}

\layoutstyle{8x11single}

\begin{document}

\title{The $X(3872)$ and other possible $XYZ$ molecular states}


\author{P.G. Ortega}{
  address={Grupo de F\'\i sica Nuclear y IUFFyM \\
            Universidad de Salamanca, 37008 Salamanca, Spain}
}

\author{J. Segovia}{
  address={Grupo de F\'\i sica Nuclear y IUFFyM \\
            Universidad de Salamanca, 37008 Salamanca, Spain}
}

\author{D.R. Entem}{
  address={Grupo de F\'\i sica Nuclear y IUFFyM \\
            Universidad de Salamanca, 37008 Salamanca, Spain}
}

\author{F. Fern\'andez}{
  address={Grupo de F\'\i sica Nuclear y IUFFyM \\
            Universidad de Salamanca, 37008 Salamanca, Spain}
}

\begin{abstract}
We perform a coupled channel calculation of the $DD^*$ and
$c\bar c$ sectors in the framework of a constituent
quark model. The interaction for the $DD^*$ states is
obtained using the Resonant Group Method (RGM) and
the underlying quark interaction model. The coupling
with the two quark system is performed using the
$^3 P_0$ model. The $X(3872)$ is found as a molecular
state with a sizable $c\bar c$ component. A comparison
with Belle and BaBar data has been done, finding a
good agreement.
Other possible molecular molecular states are discussed.
\end{abstract}

\maketitle


\section{Introduction}

In the last years a number of exciting discoveries of new hadron states, the so called 
$XYZ$ mesons, have challenged our description of the hadron spectroscopy. Among them, one of the most
mysterious states is the well established $X(3872)$. It was first discovered by
the Belle Collaboration in the $J/\psi \pi \pi$ invariant mass spectrum of the
decay $B^+\rightarrow K^+ \pi^+\pi^-J/\psi$~\citep{r1}. Its existence was soon
confirmed by BaBar~\citep{r2}, CDF~\citep{r3} and D0~\citep{r4} Collaborations.
The world average mass is $M_X=3871.2\pm 0.5\,MeV$ and its width
$\Gamma_X<2.3\,MeV$.
The measurements of the $X(3872)\rightarrow \gamma J/\psi$ decay~\citep{r5,r6}
implies an even $C$-parity. Moreover angular correlation between final
state particles in the $X(3872)\rightarrow \pi^+\pi^-J/\psi$ decay measured by
Belle~\citep{r5} suggests that the $J^{PC}=0^{++}$ and $J^{PC}=0^{+-}$ may be
ruled out and strongly favors the $J^{PC}=1^{++}$ quantum numbers although the
$2^{++}$ combination cannot be excluded. A later analysis by CDF
Collaboration~\citep{r8} of the same decay is compatible with the Belle results
and concludes from the dipion mass spectrum that the most likely quantum numbers
should be $J^{PC}=1^{++}$ but cannot totally exclude the $J^{PC}=2^{-+}$
combination .
These conclusions were confirmed by a new CDF analysis of the decay
$X(3872)\to \pi^+ \pi^- J/\psi$ followed by $J/\psi \to \mu^+ \mu^-$
excluding all the other possible quantum numbers at $99.7\,\%$
confidence level~\citep{m8}.
However the small phase space available for the decay
$X(3872)\rightarrow D^0\bar D^0\pi ^0$ observed by Belle~\citep{r9} discards the
$J=2$ leaving the $1^{++}$ assignment as the most probable option.

In the $1^{++}$ sector the only well established state in the PDG~\citep{pdg}
is the $\chi_{c_1}(1P)$ with a mass $M=3510.66\pm0.07\,MeV$. The first
excitation is expected around $3950\,MeV$. In this energy region
Belle has reported the observation of three resonant structures denoted
by $X(3940)$, $Y(3940)$ and $Z(3930)$. The last one was observed by Belle
in the $\gamma \gamma \to D\bar D$ reaction~\citep{m9a} and is already
included in the PDG as the $\chi_{c_2}(2P)$. The $X(3940)$ has been
seen as a peak in the recoiling mass spectrum of $J/\psi$ produced
in $e^+e^-$ collision. Its main decay channel is $DD^*$~\citep{m9b}.
The $Y(3940)$ appears as a threshold enhancement in the $J/\psi \omega$
invariant mass distribution of the $B\to J/\psi \omega K$ decay~\citep{m9c}.
Finally, CDF Collaboration has reported a new structure in the $B^+\to J/\psi\phi K^+$
decay at a mass of $4143\pm 2.9\pm 1.2\,MeV$~\citep{Aaltonen}. This state, called $Y(4140)$ has some 
similarities to the $Y(3940)$ state as far as decay channel is concerned.

The decays measured for the $X(3872)$ outlines a puzzling structure. In one
hand the decays into $\pi\pi J/\psi$ and $\pi\pi\pi J/\psi$ through
$\rho$ and $\omega$ mesons respectively suggests a sizable isospin
breaking, incompatible with a $c\bar c$ structure. In the other hand 
the radiative decays into $\gamma J/\psi$
and $\gamma \psi'$ suggest a sizable $c\bar c$ component.

\section{Coupled Channel calculation}

In this work we present a coupled channel calculation of the 
$1^{++}$ sector including both $c\bar c$ and $DD^*$ 
states. The calculation is done in the framework of the constituent quark 
model of Ref.~\citep{r16} widely used in hadronic spectroscopy. 

We start assuming a wave function given by
\begin{equation} \label{ec:funonda}
 | \Psi \rangle = \sum_\alpha c_\alpha | \psi_\alpha \rangle
 + \sum_\beta \chi_\beta(P) |\phi_{M_1} \phi_{M_2} \beta \rangle
\end{equation}
where $|\psi_\alpha\rangle$ are $c\bar c$ eigenstates of the two body
Hamiltonian, 
$\phi_{M_i}$ are $c\bar n$ ($\bar c n$) eigenstates describing 
the $D$ ($\bar D$) mesons, 
$|\phi_{M_1} \phi_{M_2} \beta \rangle$ is the two meson state with $\beta$ quantum
numbers coupled to total $J^{PC}$ quantum numbers
and $\chi_\beta(P)$ is the relative wave 
function between the two mesons in the molecule. 
The eigenstates
of the $C$-parity operator are given by
$DD^*\equiv D\bar D^* \pm \bar D D^*$.

We use a phenomenological $^3 P_0$ model~\citep{r21}
to couple the two and four quark systems using the
operator
\begin{eqnarray}
T&=&-3\sqrt{2}\gamma'\sum_\mu \int d^3 p d^3p' \,\delta^{(3)}(p+p')
\left[ \mathcal Y_1\left(\frac{p-p'}{2}\right) b_\mu^\dagger(p)
d_\nu^\dagger(p') \right]^{C=1,I=0,S=1,J=0}
\label{TBon}
\end{eqnarray}
where $\mu$ ($\nu=\bar \mu$) are the quark (antiquark) quantum numbers and
$\gamma'=2^{5/2} \pi^{1/2} \gamma$ and $\gamma=\frac{g}{2m}$ are dimensionless constants
that gives the strength of 
the $q\bar q$ pair creation from the vacuum. The value of gamma is fitted to
the $\psi(3770) \to DD$ decay width.

Using the wave-function from Eq. (\ref{ec:funonda}) with
the two body eigenfunctions for all the mesons we apply the
Resonant Group Method to obtain the dynamics in the
two meson sector. We finnally end up with the coupled
channel equation
\begin{equation}
	\sum_{\beta} \int \left( H^{M_1 M_2}_{\beta'\beta} (P',P) + 
	V^{eff}_{\beta'\beta} (P',P) \right)
	\chi_\beta(P) \, P^2 \, dP = E \,\chi_{\beta'}(P')
\end{equation}
where we include the $^3 S_1$ and $^3D_1$ $DD^*$ partial waves
and
\begin{equation}
	V^{eff}_{\beta'\beta}(P',P) = 
	\sum_\alpha \frac{V_{\beta'\alpha}(P') V_{\alpha\beta}(P)}{E-M_\alpha}
\end{equation}
is an effective interaction between the two mesons due to the coupling
with intermediate $c\bar c$ states with
\begin{equation}
        \langle \phi_{M_1} \phi_{M_2} \beta | T | \psi_\alpha \rangle =
        P \, V_{\beta \alpha}(P) \,\delta^{(3)}(\vec P_{\mbox{cm}}).
\end{equation}

The $c\bar c$ probabilities are given by
\begin{equation}
	c_\alpha = \frac{1}{E-M_\alpha}
	\sum_\beta\int V_{\alpha\beta} (P) \chi_\beta(P) P^2 \,dP
\end{equation}
with the normalization condition 
$1=\sum_\alpha |c_\alpha|^2 + \sum_\beta \langle \chi_\beta | \chi_\beta \rangle$.

\begin{table}
\begin{tabular}{cccccc}
\hline
 & \tablehead{1}{r}{b}{$M\,(MeV)$}
 & \tablehead{1}{r}{b}{$c\bar c(1 ^3P_1)$}
 & \tablehead{1}{r}{b}{$c\bar c(2 ^3P_1)$}
 & \tablehead{1}{r}{b}{$D^0{D^*}^0$} 
 & \tablehead{1}{r}{b}{$D^\pm{D^*}^\mp$} \\
\hline
   & 3936 & $ 0  \,\%$ & $79\,\%$   & $10.5\,\%$  & $10.5\,\%$ \\
 A & 3865 & $1\,\%$    & $32\,\%$   & $33.5\,\%$  & $33.5\,\%$ \\
   & 3467 & $95\,\%$   & $ 0  \,\%$ & $2.5\,\%$   & $2.5\,\%$  \\
\hline
   & 3937 & $   0\,\%$ & $ 79\,\%$   & $   7\,\%$  & $   14\,\%$ \\
 B & 3863 & $ 1  \,\%$ & $30\,\%$   & $46\,\%$    & $23\,\%$   \\
   & 3467 & $ 95\,\%$   & $   0\,\%$ & $  2.5\,\%$   & $  2.5\,\%$  \\
\hline
   & 3942 & $   0\,\%$ & $ 88\,\%$   & $   4\,\%$  & $   8\,\%$ \\
 C & 3871 & $ 0  \,\%$ & $ 7\,\%$   & $83\,\%$    & $10\,\%$   \\
   & 3484 & $ 97\,\%$   & $   0\,\%$ & $  1.5\,\%$   & $  1.5\,\%$  \\
\hline
\end{tabular}
\caption{\label{t1}  Masses and channel probabilities for the three
states in three different calculations. The first three states are
found when we perform and isospin symmetric calculation with a value
of $\gamma$ fit to the decay $\psi(3770)\to DD$. The second three states
shows the effect of isospin breaking in the $DD^*$ masses. The last
three states correspond to a value of $\gamma=0.19$ that fits the
experimental mass of the $X(3872)$.
The probability is shown as zero when it is less than $0.5\,\%$.}
\end{table}

The results are given in Table~\ref{t1}. Part A
corresponds to the isospin symmetric case while part B shows the
effect of isospin breaking in phase space. In both cases
no bound state is found without coupling to the $c\bar c$
sector, neither in the $I=0$ nor in the $I=1$. 
This coupling generates a new state with an energy close
to the $DD^*$ threshold.

Having in mind that the $^3 P_0$ model is probably too naive and we might be
overestimating the value of $\gamma$, we vary this parameter
to get the experimental binding energy. The probabilities are given in part C
of Table~\ref{t1}.

\section{Comparison with the data}

In order to compare the predictions of our model with the experimental
data we use a Flatt\'e-like parametrization
of the $DD^*$ amplitude following Ref.~\citep{Baru,Kala1}. 
The differential cross section to final $DD^*$ states is given by
\begin{eqnarray}
	\frac{d Br(B\to K D^0 D^{*0})}{dE} &=&
	\mathcal{B} 
	\frac{1}{2\pi} 
	\frac{\Gamma_{D^0D^{*0}}(E)}{|D(E)|^2}
\end{eqnarray}
where $\mathcal{B}$ gives the branching to $B\to KX(3872)$,
$\Gamma_{D ^0D^{*0}}(E)$ is the width calculated from the $^3 P_0$ model
and
\begin{eqnarray}
	D(E) &=& E-E_f 
	+\frac i 2(\Gamma_{D^0D^{*0}}+
	\Gamma_{D^+D^{*-}}+\Gamma(E))+\mathcal{O}(4\mu^2\epsilon/\Lambda^2).
\end{eqnarray}
where 
$\Gamma(E)$ accounts for the width due to other processes different from
the opening of the near $DD^*$ threshold.

The analysis of the $B\to KX(3872) \to K \pi^+\pi^- J/\psi$ data is more 
involved because we have to calculate the $DD^*\to \pi^+\pi^-J/\psi$
transition amplitude.

\begin{figure}
\includegraphics[height=.25\textheight]{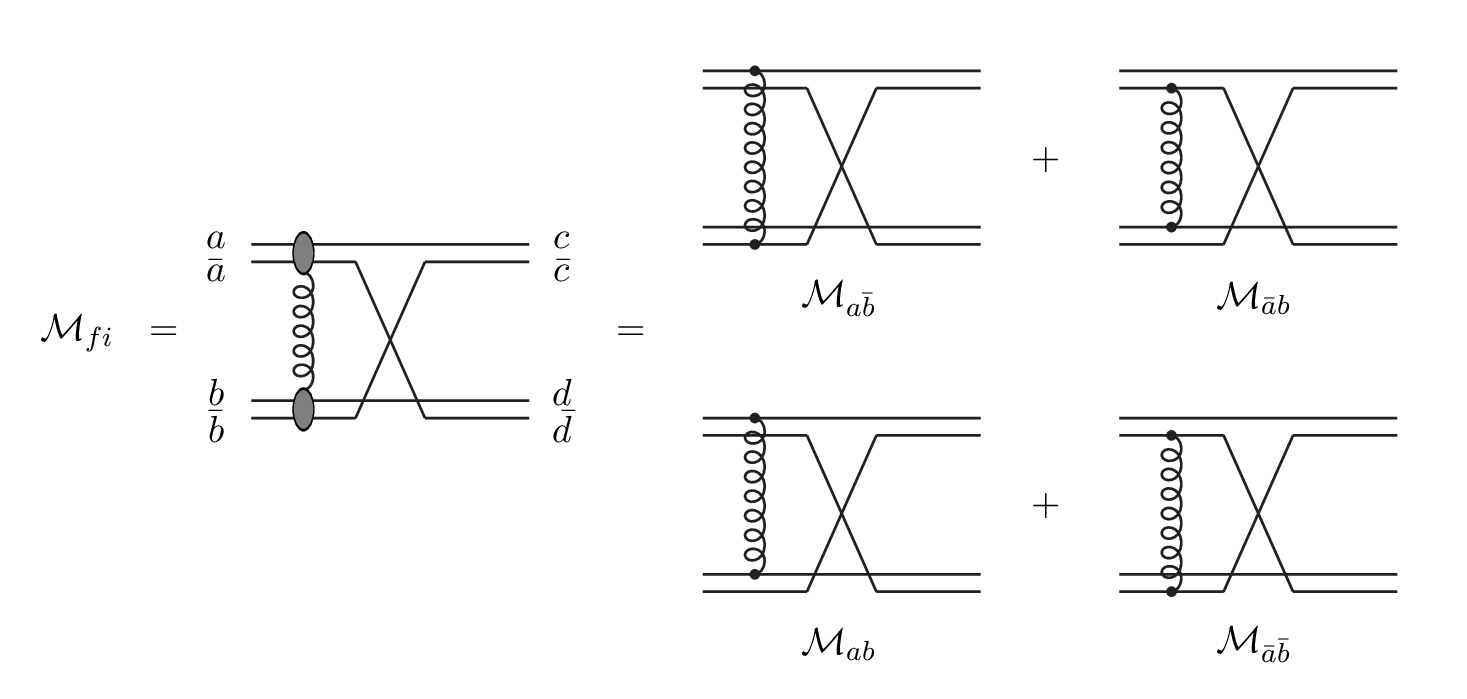}
\caption{\label{diag} Diagrams included in the quark rearrangement
process $DD^* \to \rho J/\psi$.}
\end{figure}

This can consistently be done in our formalism assuming that the process
takes place through the $DD^*$ components of the $X(3872)$ which decays
into $\rho J/\psi$ and then into the final $\pi^+\pi^-J/\psi$ state.
The decay width of the process is given by
\begin{eqnarray}
	\Gamma_{\pi^+\pi^-J/\psi} =
	\sum_{JL} \int_0^{k_{max}} dk 
	\frac{\Gamma_{\rho}}{(M_X-E_\rho-E_{J/\psi})^2+\frac{\Gamma_\rho^2}{4}}
	\left|\mathcal{M}^{JL}_{X\to \rho J/\psi}(k)\right|^2.
\end{eqnarray}
The amplitude $\mathcal{M}^{JL}_{X\to \rho J/\psi}$ is calculated
in our model by the rearrangement diagrams of Fig.~\ref{diag},
averaged with the $DD^*$ component of the $X(3872)$ wave function.
The rearrangement diagrams are calculated following Ref.~\citep{BarnesSwanson}.
The amplitude is given by
\begin{eqnarray}
	\mathcal{M}_{fi} &=& \sum_{i=a,\bar a;j=b,\bar b} \mathcal{M}_{ij}
\end{eqnarray}
where
\begin{eqnarray}
	\mathcal{M}_{ij}(\vec P',\vec P) &=& 
	\langle \phi_{M'_1} \phi_{M'_2}|H_{ij}^O| \phi_{M_1} \phi_{M_2} \rangle
	\langle \xi_{M'_1M'_2}^{SFC} |\mathcal{O}_{ij}^{SFC}| \xi_{M_1M_2}^{SFC} \rangle
\end{eqnarray}
The spin-flavor-color matrix elements are taken from
Ref.~\citep{BarnesSwanson}. 

Once the decay width $\Gamma_{\pi^+\pi^-J/\psi}$ is calculated, the 
differential rate is given by
\begin{eqnarray}
	\frac{d Br(B\to K  \pi^+\pi^-J/\psi)}{dE} &=&
	\mathcal{B} 
	\frac{1}{2\pi} \frac{\Gamma_{\pi^+\pi^-J/\psi}(E)}{|D(E)|^2}.
\end{eqnarray}
In order
to compare with the experimental data
we determine the number of events distributions from
the differential cross section following Ref.~\citep{Kala1}.
In all reactions a background is taken into account modelled
as in Ref.~\citep{Kala1}.
For the $B\to K  D^0\bar D^0\pi^0$ the $DD^{*0}$ signal
interferes with the background and so a phase $\phi^{Belle}=0^0$
and $\phi^{BaBar}=324^0$
have been introduced. Also the
experimental branching ratio $B(D^{*0}\to D^0\pi^0)=0.62$
is introduced. We use a value for $\mathcal{B}=3.5\,10^{-4}$ which
is in the order of the one used in Ref.~\citep{Kala1}.

\begin{figure}
\includegraphics[height=.3\textheight]{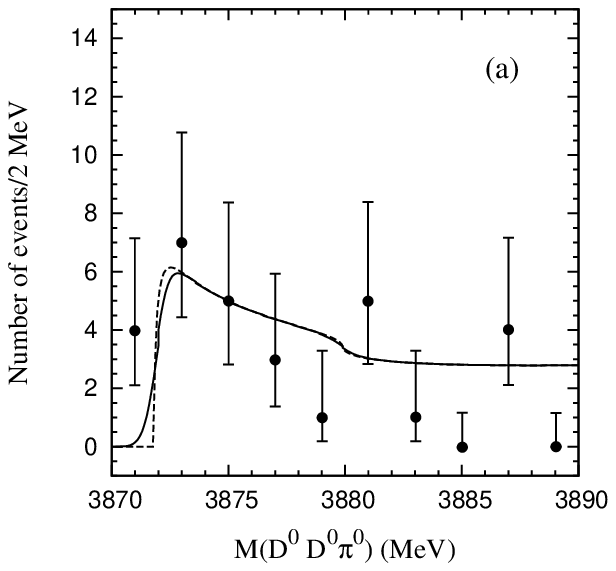}
\includegraphics[height=.3\textheight]{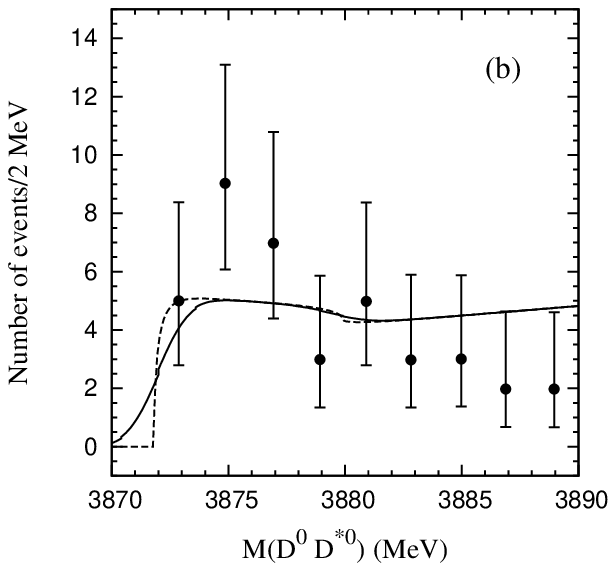}
\caption{\label{fig3} Number of events for the decay
$B\to KD^0D^0\pi^0$ measured by Belle (a) and for
the decay $B\to KD^0D^{*0}$ measured by BaBar (b).
The solid and dashed lines shows the results from
our model with and without the resolution functions
as explained in the text.}
\end{figure}

\begin{figure}
\includegraphics[height=.3\textheight]{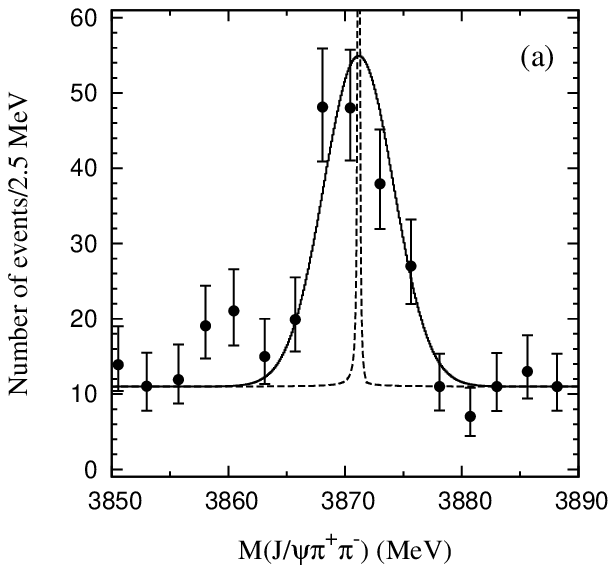}
\includegraphics[height=.3\textheight]{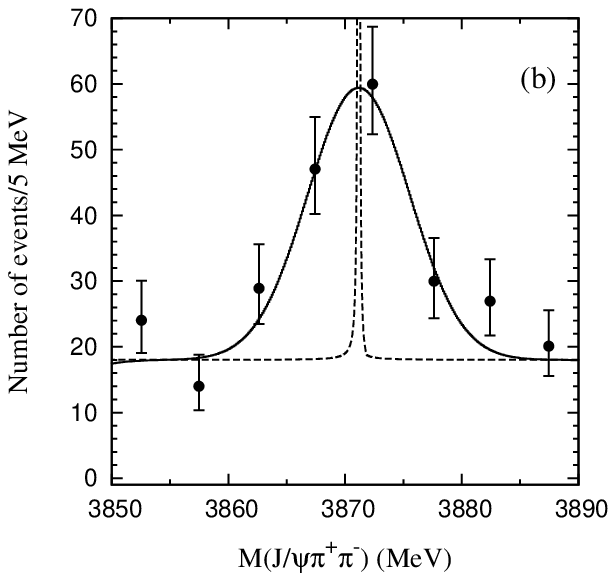}
\caption{\label{fig4} Number of events for the decay
$B\to K\pi^+\pi^-J/\psi$ measured by Belle (a) and
by BaBar (b).
The solid and dashed lines shows the results from
our model with and without the resolution functions
as explained in the text.}
\end{figure}

In Fig.~\ref{fig3} we compare our results with the
$B\to K  D^0\bar D^{0}\pi^0$ data from Belle (a) and
$B\to K  D^0\bar D^{*0}$ data from BaBar (b).
The same comparison is done in
Fig.~\ref{fig4} for the
$B\to K  \pi^+\pi^-J/\psi$ data from Belle (a)
and BaBar (b). 
In all figures the dashed lines shows
the results without resolution functions. The solid line
gives the result using the resolution functions as in
Ref.~\citep{Kala1}. All the resolution functions
are those given by Belle~\citep{BelleDDD} and BaBar~\citep{BaBarDJPsi} collaboration
with the exception of the BaBar $D^0D^{*0}$ resolution where
we use the prescription from Ref.~\citep{Kala1}.

We find a good description of the Belle $B\to K D^0 D^0\pi^0$
data
whereas the agreement is poor in the case of
the BaBar data. It is important to notice
that in the Belle analysis the mass of the
$X$ appears as $3872\,MeV$ while in the
BaBar data the resonance is located $3\,MeV$
above. The BaBar mass value does not
coincide with the
mass of the $X$ obtained in our
calculation which may be the reason
for the disagreement.

The $B\to K  \pi^+\pi^-J/\psi$ data are
equally well described for the Belle and 
BaBar experiments. In this case both
Collaborations give similar values for the mass of
the resonance, namely $3871.4\,MeV$, which are in
much better agreement with our result.

Concerning other $XYZ$ states,the $X(3940)$, decaying to $D\bar D^*$, with mass $M=3942\pm 9\,MeV$ is a good candidate to
our state with a $88\,\%$ $1^{++}$ $c\bar c$ component and mass $M=3942\,MeV$ (see Table~\ref{t1} part C).
With respect to the other two states, it has been suggested that both, the $Y(3940)$ decaying to
$J/\psi\omega$ and $Y(4140)$ decaying to $J/\psi\phi$, are $D^*D^*$ and $D^*_sD^*_s$ hadronic molecules
with $J^{PC}=0^{++}\mbox{ or }2^{++}$ respectively.
We have explore these channels but we have not found in principle any molecular bound state.


As a summary, we have shown that the $X(3872)$ emerges in a constituent
quark model calculation as a dynamically generated mixed state of a $DD^*$ molecule and 
$\chi_{c_1}(2P)$. 
This structure allows to understand simultaneously the isospin violation
showed by the experimental data and the radiative decay rates.
Furthermore, we have demonstrated that this solution explains the
new Belle data in the $D^0D^0\pi^0$ and $\pi^+\pi^-J/\psi$ decay
modes and the $\pi^+\pi^-J/\psi$ BaBar data. 
The original $\chi_{c_1}(2P)$ state acquires a significant
$DD^*$ component and can be identified with the $X(3940)$.


\begin{theacknowledgments}
This work has been partially funded by Ministerio de Ciencia y Tecnolog\'\i a
under Contract
No. FPA2007-65748, by Junta de Castilla y Le\'on under Contract No.
SA-106A07 and GR12,
by the European Community-Research Infrastructure Integrating
Activity ``Study of Strongly Interacting Matter'' (HadronPhysics2
Grant no. 227431) and
by the Spanish Ingenio-Consolider 2010 Program
CPAN (CSD2007-00042).
\end{theacknowledgments}






\begin{thebibliography}{9}

\bibitem[S. K. Choi et al. (2003)]{r1}S. K. Choi {\it et al.} (Belle Collaboration),Phys. Rev. Lett. {\bf 91}, 262001 (2003).

\bibitem[B. Aubert et al. (2005)]{r2} B. Aubert {\it et al.} (BaBar Collaboration),Phys. Rev. D {\bf 71}, 071103 (2005).

\bibitem[D. Acosta et al. (2003)]{r3} D. Acosta {\it et al.} (CDF Collaboration),Phys. Rev. Lett. {\bf 93}, 072001 (2003).

\bibitem[V. M. Abrazov et al. (2003)]{r4} V. M. Abrazov {\it et al.} (D0 Collaboration),
        Phys. Rev. Lett. {\bf 93}, 162002 (2003).

\bibitem[K. Abe et al. (2005)]{r5} K. Abe {\it et al.} (Belle Collaboration),
        arXiV:hep-ex/0505038.

\bibitem[B. Aubert et al. (2006)]{r6} B. Aubert {\it et al.} (BaBar Collaboration),
        Phys. Rev. D {\bf 74}, 071101 (2006).

\bibitem[A. Abulencia et al. (2006)]{r8} A. Abulencia {\it et al.} (CDF Collaboration),
        Phys. Rev. Lett. {\bf 96}, 102002 (2006).

\bibitem[A. Abulencia et al. (2007)]{m8} A. Abulencia {\it et al.} (CDF Collaboration),
        Phys. Rev. Lett. {\bf 98}, 132002 (2007).

\bibitem[G. Gokhroo et al. (2006)]{r9} G. Gokhroo {\it et al.} (Belle Collaboration),
        Phys. Rev. Lett. {\bf 97}, 162002 (2006).

\bibitem[C. Amsler et al. (2008)]{pdg} C. Amsler {\it et al.} (Particle Data Group),
        Phys. Lett. B {\bf 667}, 1 (2008).

\bibitem[S. Uehara et al. (2006)]{m9a} S. Uehara {\it et al.} (Belle Collaboration),
        Phys. Rev. Lett. {\bf 96}, 082003 (2006).

\bibitem[K. Abe et al. (2007)]{m9b} K. Abe {\it et al.} (Belle Collaboration),
        Phys. Rev. Lett. {\bf 98}, 082001 (2007);
        P. Pakhlov  {\it et al.} (Belle Collaboration),
        Phys. Rev. Lett. {\bf 100}, 202001 (2008);

\bibitem[S. K. Choi et al. (2005)]{m9c} S.K. Choi {\it et al.} (Belle Collaboration),
        Phys. Rev. Lett. {\bf 94}, 182002 (2005).

\bibitem[T. Aaltonen et al. ()]{Aaltonen} T. Aaltonen {\it et al.} (CDF Collaboration),
	arxiv:0903.3339.

\bibitem[J. Vijande et al. (2005)]{r16} J. Vijande, F. Fernandez and A. Valcarce,
        J. Phys. G {\bf 31}, 481 (2005).

\bibitem[L. Micu et al. (1969)]{r21} L. Micu, Nucl. Phys. B {\bf 10}, 521 (1969);
A. Le Yaouanc, L. Olivier, O. Pene and J.C. Raynal,
Phys. Rev. D {\bf 8}, 2223 (1973);
E. S. Ackleh, T. Barnes, and E. S. Swanson, Phys. Rev. D {\bf 54}, 6811 (1996).

\bibitem[V. Baru et al. (2004)]{Baru} V. Baru {\it et al.}, Phys. Lett. B {\bf 586}, 53 (2004).

\bibitem[Yu. S. Kalashnikova et al. (2009)]{Kala1} Yu.S. Kalashnikova and A.V. Nefediev,
        Phys. Rev. D {\bf 80}, 074004 (2009).

\bibitem[T. Barnes et al. (1992)]{BarnesSwanson} T. Barnes, and E.S. Swanson,
        Phys. Rev. D {\bf 46}, 131 (1992).

\bibitem[I. Adachi et al. ()]{BelleDDD} I. Adachi {\it et al.} (Belle Collaboration),
        arXiv:0810.0358.

\bibitem[B. Aubert et al. (2008)]{BaBarDJPsi} B. Aubert {\it et al.} (BaBar Collaboration),
Phys. Rev. D {\bf 77}, 111101R (2008).
\end{thebibliography}
\end{document}